\documentclass[nofootinbib,reprint,amsmath,amssymb,aps]{revtex4-2}

\usepackage{graphicx}
\usepackage{bm}
\usepackage[colorlinks]{hyperref}
\usepackage{amsthm}
\usepackage{subcaption}
\begin{document}

\title{Multi-Scale Coherence of Represented Flows}

\author{Amir Jafari}
\email{elenceq@jhu.edu}

\begin{abstract}
Many problems in nonlinear and statistical physics are formulated through represented flows,
including physical-space vector fields, phase-space drift fields, and truncated renormalization-group
beta functions. We introduce a complementary representation-dependent diagnostic for testing whether
finite-separation flow geometry is stable across observational resolution. For two separated points, states, or theories, the method compares the direction of the corresponding vector-field increment after the field has been smoothed at two resolutions. Averaging this normalized comparison over sampled separations gives a coherence matrix tied to the chosen variables, coarse graining, metric, and sampling protocol; it is a consistency test, not a coordinate-invariant quantity. We demonstrate the diagnostic in three settings. Synthetic divergence-free fields with identical Fourier amplitudes, spectra, and scalar two-point correlations nevertheless produce distinct coherence matrices, showing that second-order statistics do not determine cross-resolution increment geometry. Lorenz phase-space tests show that a smooth coordinate wrinkling changes represented drift geometry without changing the underlying dynamics, and that a weak model perturbation lowers finite-separation coherence even when local stretching proxies remain closely matched. Finally, for functional renormalization-group flows of the three-dimensional \(O(1)\) scalar theory, projected \(M=4,5,6\) LPA beta fields remain internally coherent, while cross-truncation coherence decreases as higher-order coupling directions are activated. The diagnostic provides a practical field-level check of how representations, models, and truncations preserve finite-separation flow geometry, complementing rather than replacing standard local, spectral, or fixed-point diagnostics.
\end{abstract}

\maketitle
\section{Introduction}
\label{sec:introduction}

Many calculations in nonlinear and statistical physics are formulated in terms of vector fields on a chosen space.  A velocity or magnetic field is a vector field in physical space; an autonomous dynamical system defines a drift field on phase space; and a renormalization-group or functional renormalization-group calculation defines beta functions on a finite-dimensional truncation of theory space~\cite{Kolmogorov1991,Germano1992,Lorenz1963,EckmannRuelle1985,WilsonKogut1974,Wetterich1993,BergesTetradisWetterich2002}.  In all three cases, one often needs to compare not only the local vector at a point, but also how vector-field differences between separated points are organized.

Standard diagnostics address important but different aspects of this problem.  Spectra, structure functions, and correlation functions measure amplitudes and low-order statistical structure~\cite{Kolmogorov1991}.  Lyapunov exponents, finite-time stretching rates, and local Jacobians measure local deformation and instability~\cite{EckmannRuelle1985}.  In RG and FRG calculations, fixed-point coordinates, critical exponents, and stability matrices describe local behavior near special theories~\cite{WilsonKogut1974,Wetterich1993,BergesTetradisWetterich2002}.  These quantities do not, in general, determine the directional organization of finite vector-field increments after coarse graining.  Fields with the same second-order statistics can have different phase organization; represented dynamical systems with similar local stretching can differ in finite-separation drift geometry; and nearby truncations of an FRG flow can agree near a fixed point while organizing the surrounding beta field differently.

This paper introduces a representation-dependent diagnostic for this finite-separation geometry.  Given two points separated by a displacement \(r\), we smooth the vector field at two resolutions, \(\ell\) and \(L\), form the corresponding increments across the same displacement, and compare their directions by a normalized inner product~\cite{Germano1992}.  Averaging this two-resolution cosine over base points, displacement directions, or sampled pairs gives a coherence matrix \(S_{\ell L}\).  The resolutions are taken smaller than the separation being tested, \(\ell,L<r\), so that \(r\) remains the physical, phase-space, or theory-space separation, while \(\ell\) and \(L\) specify the observational resolutions.

The diagnostic is tied to a specified representation.  Its value depends on the variables, coarse-graining prescription, metric, sampling region, and support threshold.  This is appropriate for the questions considered here: whether a chosen representation, model, truncation, or regulator preserves the finite-separation geometry of a vector field under changes of resolution.  The method is therefore a field-level consistency test for represented flows, not a coordinate-independent invariant of an abstract flow.

We first test the diagnostic in a controlled physical-space example.  We construct two two-dimensional periodic divergence-free vector fields with identical Fourier amplitudes.  Their shell spectra and scalar two-point correlations agree to numerical precision, while their phases are organized differently.  The resulting coherence matrices are distinct.  This shows directly that standard scalar second-order statistics do not determine cross-resolution increment geometry.

We then apply the diagnostic to the Lorenz vector field on a sampled phase-space region~\cite{Lorenz1963}.  This is a field-based calculation, not a trajectory diagnostic.  A smooth wrinkling of the coordinate chart is implemented by pushing forward the Lorenz vector field to the new coordinates.  The underlying dynamics is conjugate to the original Lorenz flow, but the represented drift geometry changes, and the coherence matrix responds accordingly.  A separate model-mismatch test compares the Lorenz drift field with a weakly perturbed model.  The local stretching proxies remain strongly correlated, while the true--model cross-coherence is systematically reduced.  These examples show that the diagnostic measures finite-separation structure not fixed by local linear data alone.

Finally, we apply the same construction to functional renormalization-group flows~\cite{WilsonKogut1974,Wetterich1993,BergesTetradisWetterich2002}.  In a polynomial truncation, the beta functions define a vector field on a finite-dimensional coupling space.  We study the three-dimensional \(O(1)\) scalar theory in the local potential approximation and compare projected low-order beta fields for \(M=4,5,6\) truncations.  Each projected truncation remains internally coherent under the same smoothing and sampling protocol.  Cross-truncation coherence decreases when excursions in the higher-order coupling directions are activated, with the strongest difference between the more separated \(M=4\) and \(M=6\) truncations.  The result gives a direct measure of the sensitivity of projected low-order beta-field geometry to the higher-order sector.

The paper is organized as follows.  Section~\ref{sec:increment_coherence} defines the increment-based coherence diagnostic, including metric-dependent and signed/unsigned variants.  Section~\ref{sec:synthetic_same_second_order} gives the synthetic matched-statistics example.  Section~\ref{sec:dynamical_systems} applies the diagnostic to sampled Lorenz phase-space vector fields.  Section~\ref{sec:frg_flows} applies it to FRG beta fields in theory space.  Section~\ref{sec:discussion} summarizes the interpretation and scope of the diagnostic.  Numerical details are collected in the appendices.

\section{Multiscale Coherence}
\label{sec:increment_coherence}

Let \(\mathbf{A}\) and \(\mathbf{B}\) be vector fields defined on a domain \(\mathcal{M}\).  In the applications below, \(\mathcal{M}\) may be physical space, phase space, or a finite-dimensional theory space.  Let \(\mathbf{A}_{\ell}\) and \(\mathbf{B}_{L}\) denote the corresponding fields observed at resolutions \(\ell\) and \(L\).  In a Euclidean representation we take
\begin{equation}
    \mathbf{A}_{\ell}=G_{\ell}*\mathbf{A},
    \qquad
    \mathbf{B}_{L}=G_{L}*\mathbf{B},
    \label{eq:coarse_grained_fields}
\end{equation}
where \(G_s\) is a smoothing kernel at scale \(s\).  The labels \(\ell\) and \(L\) denote observational resolutions only.  They need not be ordered in the definition.

For a displacement \(\mathbf{r}\) in the chosen representation of \(\mathcal{M}\), define the finite-resolution increments
\[    \delta_{\mathbf{r}}\mathbf{A}_{\ell}(\mathbf{x})
    =
    \mathbf{A}_{\ell}(\mathbf{x}+\mathbf{r})
    -
    \mathbf{A}_{\ell}(\mathbf{x}),
\]\[
    \delta_{\mathbf{r}}\mathbf{B}_{L}(\mathbf{x})
    =
    \mathbf{B}_{L}(\mathbf{x}+\mathbf{r})
    -
    \mathbf{B}_{L}(\mathbf{x}).\]
The local two-field, two-resolution coherence is
\begin{equation}
    S^{A,B}_{\ell L}(\mathbf{x},\mathbf{r})
    =
    \frac{
        \delta_{\mathbf{r}}\mathbf{A}_{\ell}(\mathbf{x})
        \cdot
        \delta_{\mathbf{r}}\mathbf{B}_{L}(\mathbf{x})
    }{
        \left\|\delta_{\mathbf{r}}\mathbf{A}_{\ell}(\mathbf{x})\right\|
        \left\|\delta_{\mathbf{r}}\mathbf{B}_{L}(\mathbf{x})\right\|
    } .
    \label{eq:local_increment_coherence}
\end{equation}
Thus \(S^{A,B}_{\ell L}\) is the cosine of the angle between two finite differences taken across the same displacement, but observed at two resolutions and, in general, in two fields.

The averaged statistic is
\begin{equation}
    S^{A,B}_{\ell L}(\mathbf{r})
    =
    \left\langle
    S^{A,B}_{\ell L}(\mathbf{x},\mathbf{r})
    \right\rangle_{\mathbf{x}},
    \label{eq:average_increment_coherence_vector_r}
\end{equation}
where the average may be spatial, ensemble, or sampled, depending on the application.  When directions of \(\mathbf{r}\) are averaged at fixed distance \(r=|\mathbf{r}|\), we write
\begin{equation}
    \overline{S}^{A,B}_{\ell L}(r)
    =
    \left\langle
    S^{A,B}_{\ell L}(\mathbf{x},\mathbf{r})
    \right\rangle_{\mathbf{x},|\mathbf{r}|=r}.
    \label{eq:average_increment_coherence_scalar_r}
\end{equation}
The matrix with entries \(\overline{S}^{A,B}_{\ell L}(r)\) is the coherence matrix at separation \(r\).

The observational resolutions should be smaller than the separation being probed:
\begin{equation}
    \ell<r,
    \qquad
    L<r.
    \label{eq:admissible_scales}
\end{equation}
This condition fixes the interpretation of the statistic.  One first chooses two distinct points separated by \(r\), and then asks how the measured difference between them changes as the field is viewed at different resolutions.  For this reason, the full set of admissible pairs \((\ell,L)\) is useful; one need not restrict to the triangular subset \(\ell<L\).

Where either increment is too small, Eq.~\eqref{eq:local_increment_coherence} is undefined.  Such points are omitted from the average.  Numerically, this is implemented by a support condition such as
\begin{equation}
    \left\|\delta_{\mathbf{r}}\mathbf{A}_{\ell}(\mathbf{x})\right\|
    \left\|\delta_{\mathbf{r}}\mathbf{B}_{L}(\mathbf{x})\right\|
    >
    \varepsilon .
    \label{eq:increment_support_mask}
\end{equation}

If the represented domain carries a metric \(G\), the Euclidean inner products in Eq.~\eqref{eq:local_increment_coherence} are replaced by metric inner products:
\begin{equation}
    S^{A,B}_{\ell L}(\mathbf{x},\mathbf{r})
    =
    \frac{
        \left\langle
        \delta_{\mathbf{r}}\mathbf{A}_{\ell}(\mathbf{x}),
        \delta_{\mathbf{r}}\mathbf{B}_{L}(\mathbf{x})
        \right\rangle_G
    }{
        \left\|\delta_{\mathbf{r}}\mathbf{A}_{\ell}(\mathbf{x})\right\|_G
        \left\|\delta_{\mathbf{r}}\mathbf{B}_{L}(\mathbf{x})\right\|_G
    } .
    \label{eq:metric_increment_coherence}
\end{equation}
This form is used when phase-space or theory-space coordinates are equipped with a chosen metric.  The metric is part of the protocol, not an invariant structure supplied by the diagnostic itself.

Throughout this construction, increments are understood in a specified
coordinate representation or trivialization of the vector bundle. Thus
the subtraction of vectors at separated base points in Eqs.~(2)--(7)
is not asserted to be a coordinate-free operation on an arbitrary
manifold. The chosen chart, embedding, or standardized coupling
coordinates are part of the diagnostic protocol, together with the
metric and smoothing prescription.

For comparisons across different separations it is useful to introduce the dimensionless ratios
\begin{equation}
    a=\frac{L}{r},
    \qquad
    b=\frac{\ell}{r},
    \label{eq:ratio_variables}
\end{equation}
with \(0<a<1\) and \(0<b<1\).  Averaging all admissible values falling into bins in the \((a,b)\) plane gives a compact representation of cross-resolution geometry relative to the fixed separation being probed.

The signed statistic in Eq.~\eqref{eq:local_increment_coherence} retains orientation and therefore permits cancellation between aligned and anti-aligned events.  The corresponding unsigned statistic is
\begin{equation}
    \overline{S}^{A,B,\mathrm{abs}}_{\ell L}(r)
    =
    \left\langle
    \left|
    S^{A,B}_{\ell L}(\mathbf{x},\mathbf{r})
    \right|
    \right\rangle_{\mathbf{x},|\mathbf{r}|=r}.
    \label{eq:unsigned_increment_coherence}
\end{equation}
The unsigned statistic measures the strength of local
directional organization independent of sign.  A small signed value
together with a large unsigned value indicates coherent local alignment
or anti-alignment whose sign varies over the sampled region.  In model
comparisons, the difference between signed and unsigned results helps
distinguish a loss of alignment strength from a change in sign-sensitive
cancellation structure.  For two fields defined on the same domain,
\begin{equation}
    S^{A,B}_{\ell L}(\mathbf{x},\mathbf{r})
    =
    S^{B,A}_{L\ell}(\mathbf{x},\mathbf{r}),
    \label{eq:cross_symmetry}
\end{equation}
while, in general,
\begin{equation}
    S^{A,B}_{\ell L}
    \neq
    S^{A,B}_{L\ell}
    \qquad
    (A\neq B).
    \label{eq:cross_non_symmetry}
\end{equation}
For a single field, \(A=B\), the coherence matrix is symmetric under exchange of the two resolution labels.
\begin{figure*}
\centering
\includegraphics[width=.85\textwidth]{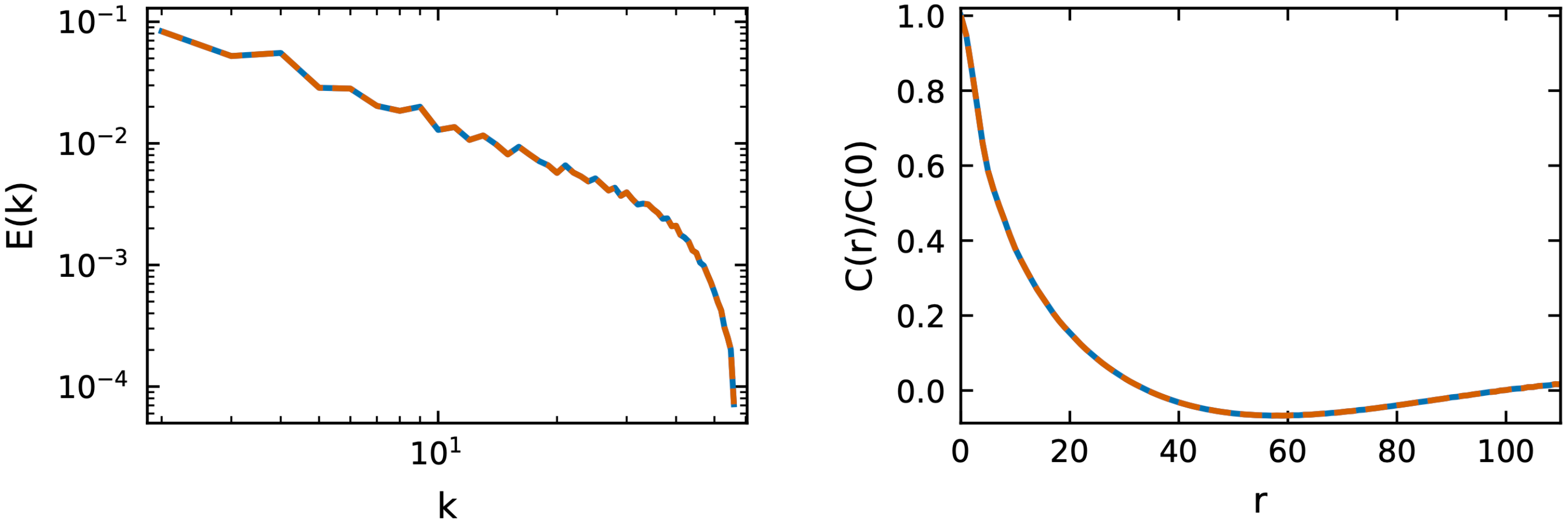}
\caption{\footnotesize
Matched second-order statistics for two synthetic divergence-free vector fields.  Solid blue denotes Field A and dashed orange denotes Field B.  The left panel shows the shell-averaged spectrum \(E(k)\), and the right panel shows the normalized scalar two-point correlation \(C(r)/C(0)\).  The two fields are indistinguishable at this second-order level, up to numerical roundoff.}
\label{fig:synthetic_second_order}
\end{figure*}

\begin{figure*}
\centering
\includegraphics[width=.85\textwidth]{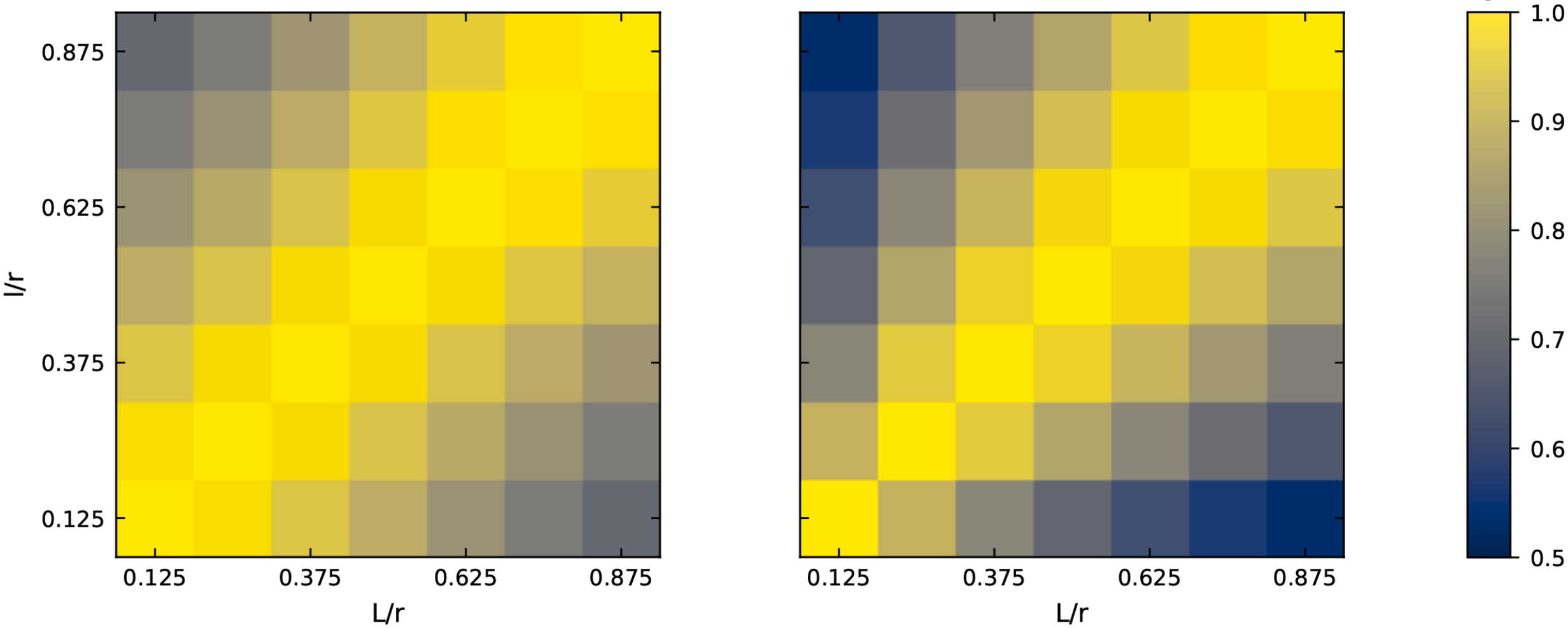}
\caption{\footnotesize
Cross-resolution coherence for the same two divergence-free fields as in Fig.~\ref{fig:synthetic_second_order}.  The left heatmap shows Field A and the right heatmap shows Field B.  The plotted quantity is the signed coherence \(S_{\ell L}\), averaged over space, separation directions, and sampled separations \(r\), then binned by the resolution ratios \(L/r\) and \(\ell/r\).  Although the two fields have identical Fourier amplitudes, matched spectra, and matched scalar two-point correlations, their coherence matrices differ, showing that second-order statistics do not determine cross-resolution directional organization.}
\label{fig:synthetic_increment_coherence}
\end{figure*}
For spatially resolved diagnostics, one may define a local windowed average around a base point \(\mathbf{x}_0\):
\begin{equation}
    S^{A,B}_{\ell L}(\mathbf{x}_0,\mathbf{r};R)
    =
    \frac{
        \int
        W_R(\mathbf{x}-\mathbf{x}_0)
        w(\mathbf{x})
        S^{A,B}_{\ell L}(\mathbf{x},\mathbf{r})
        \,\mathrm{d}\mathbf{x}
    }{
        \int
        W_R(\mathbf{x}-\mathbf{x}_0)
        w(\mathbf{x})
        \,\mathrm{d}\mathbf{x}
    } ,
    \label{eq:windowed_increment_coherence}
\end{equation}
where \(W_R\) is a window of width \(R\), and \(w\ge0\) is an optional weight emphasizing selected regions.  This version gives a local coherence field while keeping the finite-increment definition primary.

\section{Synthetic vector fields with matched second-order statistics}
\label{sec:synthetic_same_second_order}

We first test the diagnostic in a controlled setting where conventional second-order statistics are fixed by construction.  The purpose is to separate information contained in spectra and scalar two-point correlations from information contained in cross-resolution increment geometry.

The construction uses two two-dimensional periodic divergence-free vector fields on a square grid.  Both fields are generated from a scalar stream function,
\begin{equation}
    \mathbf{B}
    =
    \left(\partial_y\psi,-\partial_x\psi\right),
    \label{eq:synthetic_stream_function}
\end{equation}
so that \(\nabla\cdot\mathbf{B}=0\).  The two fields have identical Fourier amplitudes mode by mode, but different phase organization.  Field A is phase organized, while Field B has randomized phases.  Consequently, the two fields have the same spectrum and scalar two-point correlation, but need not have the same finite-separation geometry.

For each field we compute the shell-averaged spectrum
\begin{equation}
    E(k)
    =
    \frac{1}{2}
    \sum_{|\mathbf{k}'|\in k}
    \left|
    \widehat{\mathbf{B}}(\mathbf{k}')
    \right|^2 ,
    \label{eq:synthetic_spectrum}
\end{equation}
and the normalized scalar two-point correlation
\begin{equation}
    \frac{C(r)}{C(0)}
    =
    \frac{
        \left\langle
        \mathbf{B}(\mathbf{x})
        \cdot
        \mathbf{B}(\mathbf{x}+\mathbf{r})
        \right\rangle_{\mathbf{x},|\mathbf{r}|=r}
    }{
        \left\langle
        |\mathbf{B}(\mathbf{x})|^2
        \right\rangle_{\mathbf{x}}
    } .
    \label{eq:synthetic_correlation}
\end{equation}
The results are shown in Fig.~\ref{fig:synthetic_second_order}.  The two curves are indistinguishable at this level.  In the representative run, the maximum absolute spectral difference is \(6.94\times10^{-18}\), the relative difference over energetic modes is \(5.69\times10^{-16}\), and the maximum absolute difference in the normalized scalar correlation is \(1.11\times10^{-16}\).  Any difference observed below is therefore not a consequence of different Fourier amplitudes, spectra, or scalar two-point correlations.

We then compute the cross-resolution coherence.  For a coarse-grained field \(\mathbf{B}_s=G_s*\mathbf{B}\), the increment across a fixed displacement is
\begin{equation}
    \delta_{\mathbf{r}}\mathbf{B}_s(\mathbf{x})
    =
    \mathbf{B}_s(\mathbf{x}+\mathbf{r})
    -
    \mathbf{B}_s(\mathbf{x}) .
    \label{eq:synthetic_increment}
\end{equation}
The signed alignment between the increments observed at resolutions \(\ell\) and \(L\) is
\begin{equation}
    S_{\ell L}(\mathbf{x},\mathbf{r})
    =
    \frac{
        \delta_{\mathbf{r}}\mathbf{B}_{\ell}(\mathbf{x})
        \cdot
        \delta_{\mathbf{r}}\mathbf{B}_{L}(\mathbf{x})
    }{
        |\delta_{\mathbf{r}}\mathbf{B}_{\ell}(\mathbf{x})|
        |\delta_{\mathbf{r}}\mathbf{B}_{L}(\mathbf{x})|
    } .
    \label{eq:synthetic_increment_coherence_def}
\end{equation}
We average this quantity over space, over separation directions, and over several separation lengths.  The result is binned by
\begin{equation}
    a=\frac{L}{r},
    \qquad
    b=\frac{\ell}{r}.
    \label{eq:synthetic_ratio_binning}
\end{equation}
This ratio-binned representation keeps the separation \(r\) distinct from the observational resolutions \(\ell\) and \(L\).  It asks how the measured direction of the same finite difference changes when the field is viewed at different coarse-graining scales.

The resulting coherence matrices are shown in Fig.~\ref{fig:synthetic_increment_coherence}.  The two fields are no longer indistinguishable.  In the representative run, the matrix-averaged signed coherence is \(0.903\) for Field A and \(0.839\) for Field B.  The corresponding unsigned averages are \(0.934\) and \(0.898\).  Thus identical second-order statistics do not imply identical cross-resolution increment geometry.

This example establishes the basic role of the statistic.  Spectra and scalar two-point correlations constrain amplitudes and low-order correlations, but they do not determine how vector increments at different resolutions align across a fixed separation.  The coherence matrix therefore exposes geometric information that remains invisible to isotropically averaged second-order diagnostics.

\section{Dynamical systems}
\label{sec:dynamical_systems}

We next consider a standard phase-space flow.  The aim is not to define a new invariant of chaotic dynamics, but to test whether the diagnostic detects changes in the represented geometry of a drift field.  We use the Lorenz equations as a familiar example and evaluate the coherence statistic on a sampled three-dimensional region of phase space.  Thus the calculation is field based: phase-space displacements replace physical-space separations, and the averages are taken over sampled pairs of states rather than over a single trajectory.

Let \(\mathbf{f}(\mathbf{x})\) denote the Lorenz vector field.  For a coarse-grained drift field
\begin{equation}
    \mathbf{f}_{s}(\mathbf{x})
    =
    (G_s*\mathbf{f})(\mathbf{x}),
    \label{eq:ds_smoothed_field}
\end{equation}
the finite-separation increment is
\begin{equation}
    \delta_{\mathbf{r}}\mathbf{f}_{s}(\mathbf{x})
    =
    \mathbf{f}_{s}(\mathbf{x}+\mathbf{r})
    -
    \mathbf{f}_{s}(\mathbf{x}) .
    \label{eq:ds_increment}
\end{equation}
The phase-space self-coherence is then
\begin{equation}
    S_{\ell L}(\mathbf{x},\mathbf{r})
    =
    \frac{
        \delta_{\mathbf{r}}\mathbf{f}_{\ell}(\mathbf{x})
        \cdot
        \delta_{\mathbf{r}}\mathbf{f}_{L}(\mathbf{x})
    }{
        |\delta_{\mathbf{r}}\mathbf{f}_{\ell}(\mathbf{x})|
        |\delta_{\mathbf{r}}\mathbf{f}_{L}(\mathbf{x})|
    } .
    \label{eq:ds_self_coherence}
\end{equation}
A trajectory-restricted version can be useful when only time-series data are available, but that is a data-limited specialization.  Here we apply the diagnostic directly to a represented vector field on phase space.
\begin{figure*}[t]
\centering
\includegraphics[width=\textwidth]{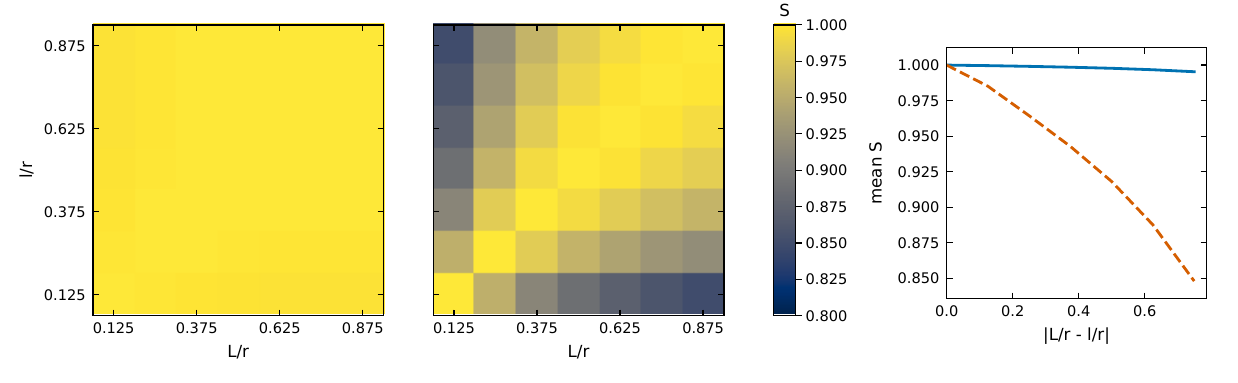}
\caption{\footnotesize
Representation dependence of cross-resolution coherence for the Lorenz vector field.  The left heatmap shows the signed coherence matrix \(S_{\ell L}\) in the original phase-space coordinates, while the middle heatmap shows the same diagnostic after a smooth wrinkling of the coordinate chart.  The right panel summarizes the same data by averaging \(S_{\ell L}\) at fixed resolution-ratio gap \(|L/r-\ell/r|\); solid blue denotes the original coordinates and dashed orange denotes the wrinkled coordinates.  The underlying dynamics is unchanged, but the represented finite-separation geometry is modified.}
\label{fig:ds_wrinkle}
\end{figure*}

The first test examines representation dependence.  We compare the standard Lorenz coordinates with a smooth wrinkled chart,
\begin{equation}
    \Phi(x,y,z)
    =
    \bigl(x,\;y+a\sin(kx),\;z\bigr).
    \label{eq:ds_wrinkle_map}
\end{equation}
The drift field in the wrinkled coordinates is the pushforward of the
original Lorenz field. If \(y=\Phi(x)\), then
\[
\widetilde f(y)
=
D\Phi(x)\,f(x)\big|_{x=\Phi^{-1}(y)} .
\]
Thus the transformed calculation compares the represented vector field
\(\widetilde f\) on the wrinkled coordinate domain, rather than merely
resampling the original components on a distorted grid.

This transformation leaves the underlying dynamical system conjugate to
the original Lorenz flow, but changes the coordinate representation in
which finite increments and smoothing are computed. The coherence change
in Fig.~\ref{fig:ds_wrinkle} is therefore a change in represented finite-separation
geometry, not a change in the abstract dynamics. In the original coordinates, the coherence matrix remains close to unity throughout the sampled range of resolution ratios.  In the wrinkled representation, the coherence decreases away from the near-diagonal region.  The same loss of coherence appears in the compressed curve obtained by averaging at fixed \(|L/r-\ell/r|\).  Quantitatively, the matrix-averaged signed coherence changes from \(0.9987\) in the original coordinates to \(0.9543\) in the wrinkled coordinates.  The corresponding off-diagonal averages change from \(0.9985\) to \(0.9467\).  Thus a smooth coordinate deformation can substantially alter cross-resolution increment geometry even though the underlying flow is unchanged.
\begin{figure*}[t]
\centering
\includegraphics[width=\textwidth]{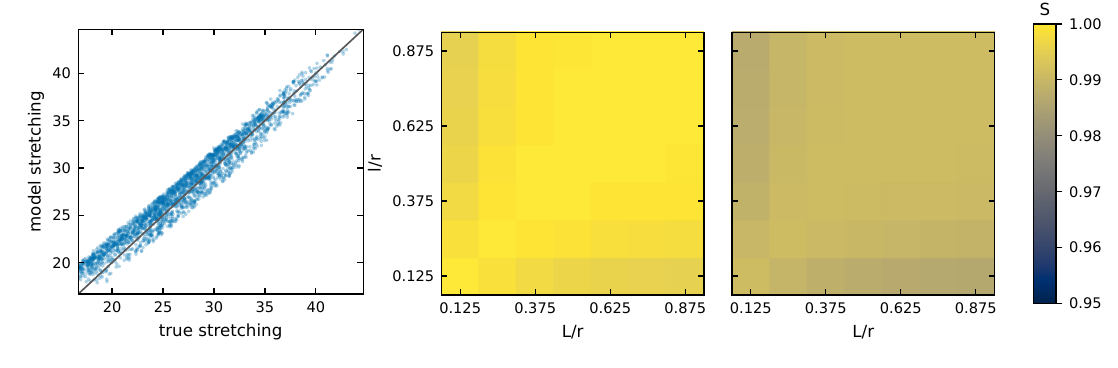}
\caption{\footnotesize
Model mismatch in a sampled Lorenz phase-space vector field.
The left panel compares the local stretching proxy
\(\sigma_{\max}(Df)\) for the Lorenz vector field with the
corresponding value for a smooth misspecified model.
The middle panel shows the true self-coherence
\(S^{f,f}_{\ell L}\), while the right panel shows the true--model
cross-coherence \(S^{f,\hat f}_{\ell L}\), computed with the same
sampling and resolution protocol.  Although the local stretching
proxies remain strongly correlated, the cross-coherence matrix is
systematically reduced relative to the true self-coherence.  Thus
agreement in a local linear proxy does not by itself imply agreement
in finite-separation, cross-resolution drift geometry.
}
\label{fig:ds_model_mismatch}
\end{figure*}

The second test considers model mismatch. We compare the Lorenz vector field
\(f\) with a smooth perturbed model field \(\hat f\) on the same sampled
phase-space region. As a conventional local comparator, we use the stretching
proxy
\[
\sigma_{\max}(Df(x)),
\]
the largest singular value of the local Jacobian. In the representative run
shown in Fig.~\ref{fig:ds_model_mismatch}, the true and model stretching proxies have Pearson
correlation \(0.979\), with median relative stretching difference \(4.3\%\).
By this local measure, the perturbed model remains close to the reference
vector field.

We then compare the true self-coherence \(S^{f,f}_{\ell L}\) with the
true--model cross-coherence \(S^{f,\hat f}_{\ell L}\). The mean signed
coherence decreases from \(0.9987\) to \(0.9898\), and the off-diagonal
average decreases from \(0.9985\) to \(0.9897\). The change is modest, as
expected for a deliberately weak perturbation, but it is systematic across the
matrix. Thus close agreement in a local stretching proxy does not by itself
imply agreement in finite-separation, cross-resolution drift geometry.

The two tests highlight complementary aspects of the diagnostic.  The coordinate-wrinkling example shows that coherence is a probe of a represented vector field rather than a coordinate-invariant dynamical quantity.  The model-mismatch example shows that local linear agreement does not determine the organization of finite drift increments across separation and resolution.  Together with the synthetic construction in Sec.~\ref{sec:synthetic_same_second_order}, these examples support the central interpretation of the coherence matrix as a measure of cross-resolution directional organization not fixed by second-order statistics or local linear data alone.

\section{Functional renormalization-group flows}
\label{sec:frg_flows}

We now apply the diagnostic to beta-function vector fields in functional renormalization-group (FRG) theory.  This application should be distinguished from an ordinary trajectory-based RG analysis.  If only a single running coupling trajectory is available, one can define a reduced direction-persistence diagnostic along that trajectory.  In a truncated FRG calculation, however, the beta functions define a vector field on a finite-dimensional theory space.  The full increment-based construction can therefore be applied with theory-space points replacing physical-space or phase-space points.

Let
\begin{equation}
    \frac{d g^i}{dt}
    =
    \beta^i(\mathbf{g}),
    \qquad
    \boldsymbol{\beta}(\mathbf{g})\in T_{\mathbf{g}}\mathcal{T},
    \label{eq:beta_flow}
\end{equation}
where \(\mathbf{g}\) denotes the couplings in a truncated theory space \(\mathcal{T}\).  The base point is a theory-space coordinate \(\mathbf{g}\), and the separation is a theory-space displacement \(\boldsymbol{\eta}\).  For two beta fields \(\boldsymbol{\beta}^{A}\) and \(\boldsymbol{\beta}^{B}\), corresponding for example to two truncations, regulators, or projections, we define
\begin{equation}
    S^{A,B}_{\ell L}(\mathbf{g},\boldsymbol{\eta})
    =
    \frac{
        \left\langle
        \delta_{\boldsymbol{\eta}}\boldsymbol{\beta}^{A}_{\ell}(\mathbf{g}),
        \delta_{\boldsymbol{\eta}}\boldsymbol{\beta}^{B}_{L}(\mathbf{g})
        \right\rangle_G
    }{
        \left\|
        \delta_{\boldsymbol{\eta}}\boldsymbol{\beta}^{A}_{\ell}(\mathbf{g})
        \right\|_G
        \left\|
        \delta_{\boldsymbol{\eta}}\boldsymbol{\beta}^{B}_{L}(\mathbf{g})
        \right\|_G
    },
    \label{eq:frg_increment_coherence}
\end{equation}
where
\begin{equation}
    \delta_{\boldsymbol{\eta}}\boldsymbol{\beta}^{A}_{\ell}(\mathbf{g})
    =
    \boldsymbol{\beta}^{A}_{\ell}(\mathbf{g}+\boldsymbol{\eta})
    -
    \boldsymbol{\beta}^{A}_{\ell}(\mathbf{g}).
    \label{eq:frg_beta_increment}
\end{equation}
Here \(\boldsymbol{\beta}_{\ell}\) denotes the beta field smoothed in theory space at resolution \(\ell\), and \(G\) is the metric used to compare tangent vectors in the sampled coupling coordinates.  In the numerical results below we use Euclidean inner products after standardizing the sampled couplings.  Other theory-space metrics can be tested by repeating the same calculation with a different \(G\).

We consider the three-dimensional \(O(1)\) scalar theory in the local potential approximation (LPA), expanded around the running minimum.  The dimensionless effective potential is truncated as
\begin{equation}
    u(\rho)
    =
    \sum_{n=2}^{M}
    \frac{\lambda_n}{n!}
    \left(\rho-\kappa\right)^n,
    \label{eq:lpa_polynomial_expansion}
\end{equation}
so that the \(M\)-th order truncation has coupling vector
\begin{equation}
    \mathbf{g}^{(M)}
    =
    \left(
    \kappa,\lambda_2,\ldots,\lambda_M
    \right).
    \label{eq:lpa_coupling_vector}
\end{equation}
The beta functions \(\boldsymbol{\beta}^{(M)}(\mathbf{g}^{(M)})\) define vector fields on the corresponding truncated theory spaces.

The question we ask is whether the low-order beta-field geometry is stable under the inclusion of higher-order couplings.  To compare truncations of different dimension, we project all beta fields onto the common low-order sector
\begin{equation}
    P_4\boldsymbol{\beta}^{(M)}
    =
    \left(
    \beta_{\kappa},
    \beta_{\lambda_2},
    \beta_{\lambda_3},
    \beta_{\lambda_4}
    \right),
    \qquad
    M=4,5,6.
    \label{eq:frg_projection}
\end{equation}
We then compute self-coherence matrices \(S^{M,M}_{\ell L}\) and cross-truncation matrices \(S^{M,N}_{\ell L}\) for the projected beta fields.  The sampled region is centered near the \(M=6\) Wilson--Fisher fixed point.  The first four coupling directions are sampled with fixed widths, while the higher-order directions \(\lambda_5\) and \(\lambda_6\) are multiplied by a width factor \(\chi\).  Thus \(\chi=0\) suppresses excursions in the higher-order sector, whereas increasing \(\chi\) activates the dependence of the projected low-order flow on higher-order couplings.

\begin{figure*}[t]
\centering
\includegraphics[width=\textwidth]{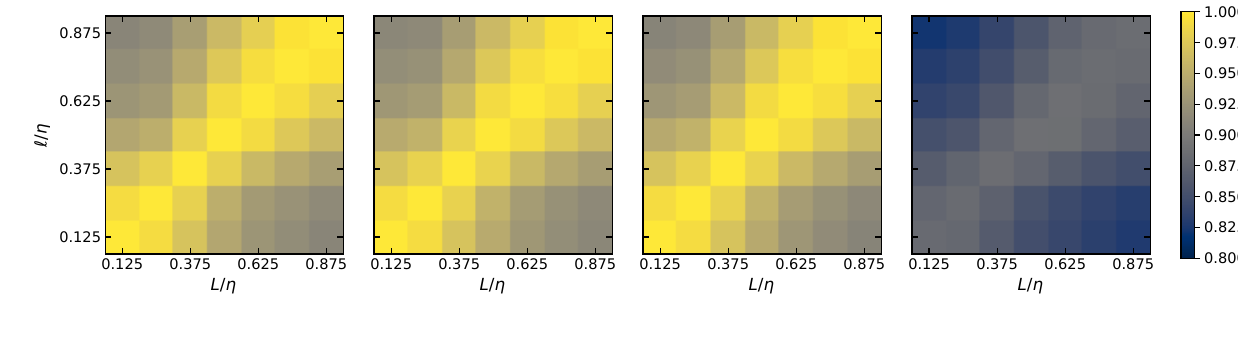}
\caption{\footnotesize
FRG truncation geometry in the three-dimensional \(O(1)\) LPA flow.  From left to right, the heatmaps show the projected self-coherence matrices for \(M=4\), \(M=5\), and \(M=6\) polynomial truncations, followed by the projected \(M=4\)--\(M=6\) cross-coherence matrix.  The plotted quantity is the signed increment coherence \(S_{\ell L}\), averaged over sampled theory-space points, displacement directions, and displacement lengths, and binned by \(L/\eta\) and \(\ell/\eta\).  The self-coherence matrices remain high and structurally similar, while the cross-truncation matrix is systematically reduced.}
\label{fig:frg_truncation_coherence}
\end{figure*}

Figure~\ref{fig:frg_truncation_coherence} shows the coherence matrices at \(\chi=1\).  The first three heatmaps show the projected self-coherence of the \(M=4\), \(M=5\), and \(M=6\) beta fields.  These matrices remain highly coherent and have similar structure across the \((L/\eta,\ell/\eta)\) plane, where \(\eta=\|\boldsymbol{\eta}\|_G\).  The fourth heatmap shows the projected \(M=4\)--\(M=6\) cross-truncation coherence.  This matrix is systematically lower, indicating that the finite-separation geometry of the projected beta field is not preserved when the higher-order sector is activated.

\begin{figure*}[t]
\centering
\includegraphics[width=\textwidth]{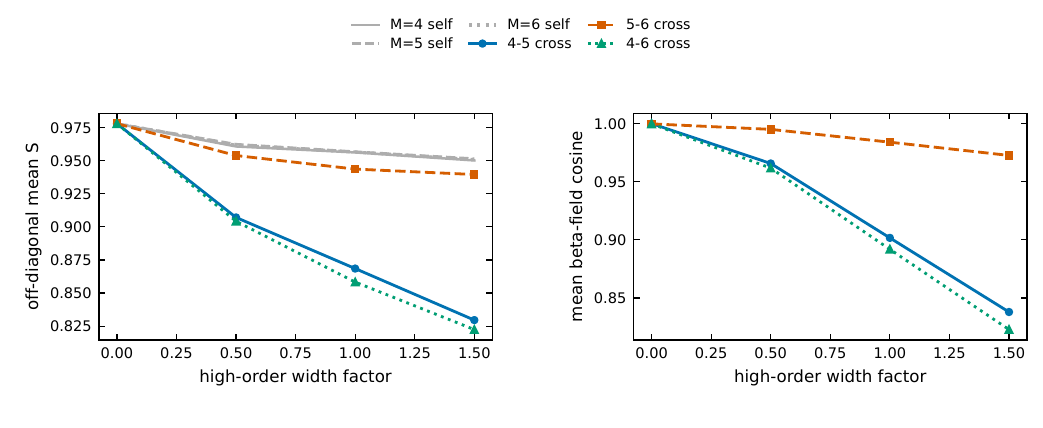}
\caption{\footnotesize
Truncation dependence of projected beta-field geometry.  Left: off-diagonal mean increment coherence as a function of the width factor \(\chi\) controlling excursions in the higher-order coupling sector.  The self-coherences of the \(M=4\), \(M=5\), and \(M=6\) truncations remain high, while the cross-truncation coherences decrease as higher-order couplings are activated.  Right: mean pointwise beta-field cosine for the same projected truncation pairs.  The pointwise comparison also detects truncation mismatch, but the coherence statistic probes finite-separation, cross-resolution beta-field geometry rather than only pointwise beta-vector alignment.}
\label{fig:frg_truncation_scan}
\end{figure*}

The same effect is quantified in Fig.~\ref{fig:frg_truncation_scan}.  We define the off-diagonal mean coherence
\begin{equation}
    \mathcal{C}_{MN}(\chi)
    =
    \left\langle
    S^{M,N}_{\ell L}
    \right\rangle_{\ell\ne L,\mathbf{g},\boldsymbol{\eta}},
    \label{eq:frg_offdiag_mean}
\end{equation}
where the average is over off-diagonal resolution pairs and over the sampled theory-space displacements.  At \(\chi=0\), all projected truncations agree closely: the self- and cross-coherences are all approximately \(0.978\).  As the higher-order sector is activated, the self-coherences decrease only mildly and remain close to \(0.95\).  By contrast, the cross-truncation coherences separate strongly.  At \(\chi=1\), the self-coherences are
\begin{equation}
    \mathcal{C}_{44}=0.956,
    \qquad
    \mathcal{C}_{55}=0.957,
    \qquad
    \mathcal{C}_{66}=0.956,
    \label{eq:frg_self_values}
\end{equation}
whereas the cross-truncation values are
\begin{equation}
    \mathcal{C}_{45}=0.869,
    \qquad
    \mathcal{C}_{56}=0.944,
    \qquad
    \mathcal{C}_{46}=0.858.
    \label{eq:frg_cross_values}
\end{equation}
At \(\chi=1.5\), the \(M=4\)--\(M=6\) cross-coherence decreases further to \(0.822\), while all three self-coherences remain near \(0.95\).  The ordering is physically natural: neighboring truncations \(M=5\) and \(M=6\) remain closest, whereas the more separated \(M=4\) and \(M=6\) truncations show the largest geometric mismatch. The pointwise beta-vector cosine in Fig.~6 shows a similar truncation
hierarchy; the coherence statistic should therefore be read here as the
finite-separation, cross-resolution counterpart of this local comparison.

These results show that the projected low-order beta-field geometry is
sensitive to excursions in the higher-order sector. Each truncation
remains internally coherent under the same smoothing and sampling
protocol, but the projected finite-separation increments are not
organized identically when the dependence on \(\lambda_5\) and
\(\lambda_6\) is activated. The effect is ordered by truncation
distance: the neighboring \(M=5\) and \(M=6\) truncations remain closest,
whereas the projected \(M=4\) and \(M=6\) beta fields show the largest
difference.

This comparison is complementary to fixed-point and stability-matrix
analyses, rather than a substitute for them.  Fixed-point coordinates and critical exponents describe local behavior near special points in theory space.  The present diagnostic instead probes an extended sampled region and asks whether beta-field differences across finite coupling-space separations are preserved across resolution and truncation.  In this sense, cross-resolution coherence provides a field-level geometric comparison of FRG truncations, rather than another estimate of local critical data.

\section{Discussion}
\label{sec:discussion}

The results above support a common interpretation: cross-resolution coherence measures a component of flow geometry that is not fixed by standard spectral, local, or fixed-point diagnostics.  The statistic asks whether the direction of a finite difference across a fixed separation remains stable when the underlying flow is observed at different resolutions.  This is a narrower question than determining a spectrum, correlation function, local Jacobian, Lyapunov exponent, fixed point, or critical exponent.  It is also a distinct question.

The synthetic example isolates this distinction in the most controlled setting.  The two divergence-free fields have identical Fourier amplitudes, indistinguishable shell spectra, and indistinguishable scalar two-point correlations, yet their coherence matrices differ.  The difference therefore cannot be attributed to conventional second-order statistics.  It comes from phase organization and from the way vector increments at different resolutions remain, or fail to remain, directionally aligned across the same separation.  Thus the coherence matrix detects cross-resolution geometric organization that can survive after amplitude-based second-order information has been fixed.

The dynamical-systems examples clarify the role of representation.  The coordinate-wrinkling test changes the represented Lorenz vector field without changing the underlying dynamics.  The coherence matrix responds because the diagnostic is not a coordinate-invariant quantity of the Lorenz flow; it is a probe of the represented drift geometry.  The model-mismatch test makes a complementary point.  A weakly perturbed model can closely match a local stretching proxy while still showing a systematic reduction in true--model cross-coherence.  Local linear information therefore does not by itself determine finite-separation drift geometry.

The FRG example shows that the same idea is meaningful in theory space.  In a truncated FRG calculation, beta functions define vector fields on a finite-dimensional coupling space.  The coherence diagnostic can therefore compare the geometry of beta-field increments across truncations.  In the three-dimensional \(O(1)\) LPA test, the projected \(M=4\), \(M=5\), and \(M=6\) beta fields remain internally coherent, but their cross-truncation coherences separate when higher-order coupling directions are activated.  The hierarchy is physically natural: neighboring truncations remain closer than more widely separated truncations.  This suggests that cross-resolution coherence can serve as a field-level comparison of truncations, complementary to fixed-point coordinates, critical exponents, and stability matrices.

The statistic is protocol dependent: its value depends on the chosen
variables, smoothing kernel, metric, sampling region, displacement
distribution, and support threshold.  It is therefore most useful in
comparisons where this protocol is fixed.  For representation
assessment and model comparison, this dependence is useful: the question
is precisely whether a chosen representation, model, truncation, or
regulator reproduces the same finite-separation geometry under the same
diagnostic protocol.

Several extensions are natural.  The unsigned statistic can separate the strength of local alignment from sign-sensitive cancellation.  Local windowed averages can identify regions of physical space, phase space, or theory space where coherence is unusually high or low.  Alternative metrics can be compared in theory-space applications.  Bootstrap and surrogate tests can be used to separate robust geometric organization from sampling artifacts.  In applications where only a single RG trajectory or time series is available, one may define a trajectory-restricted direction-persistence diagnostic, but this should be distinguished from the full field-based construction used in the main FRG example.

The central use of the diagnostic is comparative.
Cross-resolution coherence does not replace spectra, structure
functions, Lyapunov exponents, local Jacobians, fixed-point data, or
critical exponents.  It complements them by measuring whether finite
flow differences retain their direction across observational resolution.
The examples above show that this quantity can distinguish geometric
organization of represented flows that is not determined by standard
local, spectral, or fixed-point diagnostics.

\appendix
\section{Numerical construction of the synthetic vector fields}
\label{app:synthetic_numerics}

This appendix gives the numerical details for the synthetic example in Sec.~\ref{sec:synthetic_same_second_order}.  The construction is designed to compare two periodic divergence-free vector fields whose conventional second-order statistics are identical by construction, while their phase organization, and hence their finite-separation geometry, is different.

The fields are defined on a two-dimensional periodic grid with \(N=256\) points in each direction.  Each field is generated from a scalar stream function,
\begin{equation}
    \mathbf{B}(\mathbf{x})
    =
    \bigl(\partial_y\psi(\mathbf{x}),-\partial_x\psi(\mathbf{x})\bigr),
\end{equation}
which enforces \(\nabla\cdot\mathbf{B}=0\) to spectral accuracy.  The Fourier coefficients of the stream function are written as
\begin{equation}
    \widehat{\psi}(\mathbf{k})
    =
    A(k)\exp\!\left(i\varphi_{\mathbf{k}}\right),
    \qquad
    k=|\mathbf{k}|.
\end{equation}
Both fields use the same modal amplitude,
\begin{equation}
    A(k)
    =
    k^{-p}
    \exp\!\left[
        -\frac{1}{2}
        \left(\frac{k}{k_{\rm damp}}\right)^6
    \right],
\end{equation}
for \(2\le k\le 56\), with \(A(k)=0\) outside this interval.  The parameters are
\begin{equation}
    p=2.1,
    \qquad
    k_{\rm damp}=48.
\end{equation}
The zero mode is omitted, and Hermitian symmetry,
\begin{equation}
    \widehat{\psi}(-\mathbf{k})
    =
    \widehat{\psi}(\mathbf{k})^*,
\end{equation}
is imposed so that the stream function is real.

The two fields differ only in their phases.  Field A uses the organized phase prescription
\begin{equation}
    \varphi_{\mathbf{k}}^{A}
    =
    -\frac{2\pi}{N}(k_x x_0+k_y y_0)
    +0.35\sin(3\theta_{\mathbf{k}})
    +0.15\cos(5\theta_{\mathbf{k}}),
\end{equation}
where
\begin{equation}
    \theta_{\mathbf{k}}=\mathrm{atan2}(k_y,k_x),
    \qquad
    x_0=0.47N,
    \qquad
    y_0=0.53N.
\end{equation}
Field B uses phases drawn uniformly from \([0,2\pi)\), subject to the same Hermitian symmetry condition.  The pseudorandom seed is \(20260512\).  Since the amplitude \(A(k)\) is identical mode by mode, the two fields have the same Fourier amplitudes; only the phase organization differs.  Both fields are normalized by the same rms convention,
\begin{equation}
    \left\langle |\mathbf{B}^{A}|^2\right\rangle^{1/2}
    =
    \left\langle |\mathbf{B}^{B}|^2\right\rangle^{1/2}
    =
    1.
\end{equation}
The maximum measured divergence errors are
\begin{equation}
    \max|\nabla\cdot\mathbf{B}^{A}|=3.25\times 10^{-13},
    \qquad
    \max|\nabla\cdot\mathbf{B}^{B}|=9.30\times 10^{-14}.
\end{equation}

The shell-averaged spectrum is computed as
\begin{equation}
    E(k)
    =
    \frac{1}{2}
    \sum_{|\mathbf{k}'|\in k}
    \left(
        |\widehat{B}_x(\mathbf{k}')|^2
        +
        |\widehat{B}_y(\mathbf{k}')|^2
    \right),
\end{equation}
with modes assigned to integer shells by rounding \(|\mathbf{k}'|\).  The scalar two-point correlation is
\begin{equation}
    C(\mathbf{r})
    =
    \left\langle
        \mathbf{B}(\mathbf{x})\cdot
        \mathbf{B}(\mathbf{x}+\mathbf{r})
    \right\rangle_{\mathbf{x}},
\end{equation}
which is radially averaged over periodic displacement vectors and normalized by \(C(0)\).  The two fields agree in these diagnostics to roundoff accuracy.  The maximum absolute spectral difference is
\begin{equation}
    \max_k |E_A(k)-E_B(k)|
    =
    6.94\times 10^{-18}.
\end{equation}
Restricting to energetic shells with \(E_A(k)>10^{-8}E_{A,\max}\), the maximum relative spectral difference is
\begin{equation}
    \max
    \frac{|E_A(k)-E_B(k)|}{E_A(k)}
    =
    5.69\times 10^{-16}.
\end{equation}
The maximum absolute difference in the normalized scalar correlation is
\begin{equation}
    \max_r
    \left|
    \frac{C_A(r)}{C_A(0)}
    -
    \frac{C_B(r)}{C_B(0)}
    \right|
    =
    1.11\times 10^{-16}.
\end{equation}

The coherence calculation uses Gaussian-filtered fields,
\begin{equation}
    \mathbf{B}_s=G_s*\mathbf{B}.
\end{equation}
For a separation vector \(\mathbf{r}\), the filtered increment is
\begin{equation}
    \delta_{\mathbf{r}}\mathbf{B}_{s}(\mathbf{x})
    =
    \mathbf{B}_{s}(\mathbf{x}+\mathbf{r})
    -
    \mathbf{B}_{s}(\mathbf{x}).
\end{equation}
The signed cross-resolution alignment is
\begin{equation}
    S_{\ell L}(\mathbf{x},\mathbf{r})
    =
    \frac{
        \delta_{\mathbf{r}}\mathbf{B}_{\ell}(\mathbf{x})
        \cdot
        \delta_{\mathbf{r}}\mathbf{B}_{L}(\mathbf{x})
    }{
        |\delta_{\mathbf{r}}\mathbf{B}_{\ell}(\mathbf{x})|\,
        |\delta_{\mathbf{r}}\mathbf{B}_{L}(\mathbf{x})|
    }.
\end{equation}
Points for which the denominator is smaller than \(10^{-14}\) are omitted.

The sampled separation lengths are
\begin{equation}
    r\in\{32,48,64,80,96\},
\end{equation}
in grid units.  For each separation, the resolution ratios are
\begin{equation}
    \frac{\ell}{r},\frac{L}{r}
    \in
    \{0.125,0.25,0.375,0.50,0.625,0.75,0.875\}.
\end{equation}
The separation directions are
\begin{equation}
    (\pm r,0),
    \qquad
    (0,\pm r),
    \qquad
    (\pm d,\pm d),
    \qquad
    d=\mathrm{round}(r/\sqrt{2}),
\end{equation}
with all four diagonal sign choices included.  The coherence matrices are averaged over the full periodic grid, over the sampled directions, and over the sampled values of \(r\).  The final matrices are displayed as functions of \(L/r\) and \(\ell/r\).

The matrix-averaged signed coherences are
\begin{equation}
    \langle S\rangle_A=0.903480,
    \qquad
    \langle S\rangle_B=0.838675.
\end{equation}
The corresponding unsigned averages are
\begin{equation}
    \langle |S|\rangle_A=0.934427,
    \qquad
    \langle |S|\rangle_B=0.898421.
\end{equation}
Thus the two fields are indistinguishable by the spectrum and scalar two-point correlation, while their cross-resolution increment coherence differs measurably.

\section{Numerical details for the FRG calculation}
\label{app:frg_numerics}

This appendix gives the numerical details for the FRG calculation in Sec.~\ref{sec:frg_flows}.  The example uses the three-dimensional \(O(1)\) scalar theory in the local potential approximation.  The dimensionless effective potential is expanded about the running minimum,
\begin{equation}
    u(\rho)
    =
    \sum_{n=2}^{M}
    \frac{\lambda_n}{n!}
    (\rho-\kappa)^n,
\end{equation}
with coupling vector
\begin{equation}
    \mathbf{g}^{(M)}
    =
    (\kappa,\lambda_2,\ldots,\lambda_M).
\end{equation}
We use polynomial truncations \(M=4,5,6\).

The normalized LPA flow used in the numerical test is
\begin{equation}
    \partial_t u
    =
    -d\,u
    +(d-2)\rho u'
    +\ell_0\!\left(u'+2\rho u''\right),
    \qquad
    d=3,
\end{equation}
with the Litim-like analytic threshold form
\begin{equation}
    \ell_0(w)=\frac{1}{1+w}.
\end{equation}
The overall threshold prefactor is absorbed into the coupling normalization and is kept fixed throughout the calculation.  Defining the right-hand side of the flow equation as \(F(\rho)\), the running-minimum condition \(u'(\kappa)=0\) gives
\begin{equation}
    \beta_{\kappa}
    =
    -\frac{F'(\kappa)}{\lambda_2}.
\end{equation}
For \(n\ge 2\), the beta functions for the polynomial couplings are
\begin{equation}
    \beta_{\lambda_n}
    =
    F^{(n)}(\kappa)
    +
    \beta_{\kappa}\lambda_{n+1},
    \qquad
    \lambda_{M+1}=0.
\end{equation}
The derivatives \(F^{(n)}(\kappa)\) are obtained by evaluating \(F(\kappa+x)\) on a symmetric stencil
\begin{equation}
    x\in[-0.16,0.16],
\end{equation}
and fitting a local polynomial of order \(M\).  The number of stencil points is
\begin{equation}
    N_{\rm stencil}=\max(2M+5,13).
\end{equation}

The fixed points used to define the sampled region are obtained by solving \(\boldsymbol{\beta}^{(M)}=0\).  The fixed points are
\begin{align}
\mathbf{g}^{(4)}_* &=
\begin{pmatrix}
1.8121383\\
0.12681138\\
0.030113132\\
0.006607595
\end{pmatrix},
\\[0.5em]
\mathbf{g}^{(5)}_* &=
\begin{pmatrix}
1.8094707\\
0.12714002\\
0.030131198\\
0.0066393232\\
-0.00023311741
\end{pmatrix},
\\[0.5em]
\mathbf{g}^{(6)}_* &=
\begin{pmatrix}
1.8154559\\
0.12607025\\
0.029787713\\
0.0062302719\\
-0.00027105714\\
-0.001219276
\end{pmatrix}.
\end{align}
The corresponding beta-function residuals are
\begin{equation}
    1.115\times10^{-11},
    \qquad
    2.189\times10^{-10},
    \qquad
    2.232\times10^{-10},
\end{equation}
for \(M=4,5,6\), respectively.  The largest real parts of the stability eigenvalues in this normalization are
\begin{equation}
    3.645966,
    \qquad
    6.581355,
    \qquad
    10.139414.
\end{equation}
These fixed-point data are used to define the sampled region and local comparison scales; they are not intended as precision estimates of critical exponents.

The theory-space sampling is centered at the \(M=6\) fixed point.  A six-dimensional standardized coordinate \(\mathbf{X}\in[-1,1]^6\) is sampled uniformly using pseudorandom seed \(20260513\).  The first four coupling directions are sampled with widths
\begin{equation}
    (0.70,\;0.110,\;0.070,\;0.030),
\end{equation}
corresponding to \((\kappa,\lambda_2,\lambda_3,\lambda_4)\).  The higher-order directions have base widths
\begin{equation}
    (0.040,\;0.022),
\end{equation}
corresponding to \((\lambda_5,\lambda_6)\), and are multiplied by a width factor
\begin{equation}
    \chi\in\{0,0.5,1.0,1.5\}.
\end{equation}
Thus \(\chi=0\) fixes the higher-order couplings at their central values, while increasing \(\chi\) samples progressively larger excursions in the higher-order sector.  For each \(\chi\), \(10000\) valid sampled points are retained.  A point is retained only if the \(M=4\), \(M=5\), and \(M=6\) beta fields are all finite.

The beta fields are compared in the common projected low-order sector
\begin{equation}
    P_4\boldsymbol{\beta}^{(M)}
    =
    \left(
    \beta_{\kappa},
    \beta_{\lambda_2},
    \beta_{\lambda_3},
    \beta_{\lambda_4}
    \right),
    \qquad
    M=4,5,6.
\end{equation}
Before computing distances and inner products, the projected beta components are scaled by the same four widths used for the corresponding coupling directions.  This is equivalent to using a Euclidean metric in the standardized low-order coordinates.

Coarse graining of the projected beta fields is performed on the sampled theory-space cloud.  For a query point \(\mathbf{X}\), the smoothed projected beta field at resolution \(s\) is
\begin{equation}
    \boldsymbol{\beta}^{(M)}_s(\mathbf{X})
    =
    \frac{
        \sum_{j\in{\cal N}_K(\mathbf{X})}
        \exp\!\left[-|\mathbf{X}-\mathbf{X}_j|^2/(2s^2)\right]
        P_4\boldsymbol{\beta}^{(M)}(\mathbf{X}_j)
    }{
        \sum_{j\in{\cal N}_K(\mathbf{X})}
        \exp\!\left[-|\mathbf{X}-\mathbf{X}_j|^2/(2s^2)\right]
    },
\end{equation}
with \(K=170\) nearest neighbors.  Theory-space pairs are sampled with standardized separation
\begin{equation}
    \eta\in\{0.35,0.55,0.75\},
\end{equation}
and tolerance
\begin{equation}
    |\mathbf{X}_j-\mathbf{X}_i|
    \in
    [(1-\tau)\eta,(1+\tau)\eta],
    \qquad
    \tau=0.18.
\end{equation}
For each \(\eta\), \(650\) pairs are sampled.  The resolution ratios are
\begin{equation}
    \frac{\ell}{\eta},\frac{L}{\eta}
    \in
    \{0.125,0.25,0.375,0.50,0.625,0.75,0.875\}.
\end{equation}
For a given ratio \(q\), the smoothing bandwidth is \(s=q\eta\), with lower cutoff \(s=0.025\) and upper cutoff \(s=0.98\eta\).  The denominator threshold for the coherence calculation is \(10^{-14}\).

For truncations \(M,N\in\{4,5,6\}\), the signed projected coherence is
\begin{equation}
    S^{M,N}_{\ell L}(\mathbf{X}_i,\mathbf{X}_j)
    =
    \frac{
        \delta_{\boldsymbol{\eta}}\boldsymbol{\beta}^{(M)}_{\ell}(\mathbf{X}_i)
        \cdot
        \delta_{\boldsymbol{\eta}}\boldsymbol{\beta}^{(N)}_{L}(\mathbf{X}_i)
    }{
        \left|
        \delta_{\boldsymbol{\eta}}\boldsymbol{\beta}^{(M)}_{\ell}(\mathbf{X}_i)
        \right|
        \left|
        \delta_{\boldsymbol{\eta}}\boldsymbol{\beta}^{(N)}_{L}(\mathbf{X}_i)
        \right|
    },
\end{equation}
where
\begin{equation}
    \delta_{\boldsymbol{\eta}}\boldsymbol{\beta}^{(M)}_{s}(\mathbf{X}_i)
    =
    \boldsymbol{\beta}^{(M)}_{s}(\mathbf{X}_j)
    -
    \boldsymbol{\beta}^{(M)}_{s}(\mathbf{X}_i).
\end{equation}
The off-diagonal mean coherence used in Fig.~\ref{fig:frg_truncation_scan} is
\begin{equation}
    \mathcal{C}_{MN}(\chi)
    =
    \left\langle
    S^{M,N}_{\ell L}
    \right\rangle_{\ell\ne L,\mathbf{X},\boldsymbol{\eta}}.
\end{equation}

The numerical values of \(\mathcal{C}_{MN}\) are listed in Table~\ref{tab:frg_coherence_values}.  At \(\chi=0\), all projected fields agree closely.  As the higher-order sector is activated, the self-coherences remain high, while the cross-truncation coherences decrease, with the largest mismatch between \(M=4\) and \(M=6\).

\begin{table}[t]
\caption{\footnotesize
Off-diagonal mean projected coherence values for the FRG truncation test.  The self-coherences remain high as the higher-order sector is activated, while cross-truncation coherences decrease, especially for the more separated \(M=4\) and \(M=6\) truncations.}
\label{tab:frg_coherence_values}
\begin{ruledtabular}
\begin{tabular}{c|cccccc}
\(\chi\) &
\(\mathcal{C}_{44}\) &
\(\mathcal{C}_{55}\) &
\(\mathcal{C}_{66}\) &
\(\mathcal{C}_{45}\) &
\(\mathcal{C}_{56}\) &
\(\mathcal{C}_{46}\) \\
\hline
0.0 & 0.977916 & 0.977917 & 0.977945 & 0.977903 & 0.977907 & 0.977880 \\
0.5 & 0.960678 & 0.962407 & 0.961578 & 0.907094 & 0.953828 & 0.904238 \\
1.0 & 0.956160 & 0.956774 & 0.956460 & 0.868516 & 0.943602 & 0.858313 \\
1.5 & 0.950116 & 0.951507 & 0.950743 & 0.829626 & 0.939504 & 0.822478
\end{tabular}
\end{ruledtabular}
\end{table}

For comparison with a pointwise diagnostic, the mean cosine between the projected beta vectors themselves is also computed.  The values are given in Table~\ref{tab:frg_beta_cosine_values}.  This comparison detects truncation mismatch, but unlike the coherence statistic it does not probe finite-separation, cross-resolution beta-field geometry.

\begin{table}[t]
\caption{\footnotesize
Mean pointwise cosine between projected beta vectors for the FRG truncation test.}
\label{tab:frg_beta_cosine_values}
\begin{ruledtabular}
\begin{tabular}{c|ccc}
\(\chi\) &
\(M=4\) vs. \(M=5\) &
\(M=5\) vs. \(M=6\) &
\(M=4\) vs. \(M=6\) \\
\hline
0.0 & 0.999973 & 0.999743 & 0.999722 \\
0.5 & 0.965658 & 0.995002 & 0.961675 \\
1.0 & 0.901682 & 0.983978 & 0.891847 \\
1.5 & 0.837934 & 0.972585 & 0.822819
\end{tabular}
\end{ruledtabular}
\end{table}

\bibliography{DSFRG}
\end{document}